\documentclass[twocolumn,prb,showpacs,superscriptaddress]{revtex4}
\AtBeginDocument{
\heavyrulewidth=.08em
\lightrulewidth=.05em
\cmidrulewidth=.03em
\belowrulesep=.65ex
\belowbottomsep=0pt
\aboverulesep=.4ex
\abovetopsep=0pt
\cmidrulesep=\doublerulesep
\cmidrulekern=.5em
\defaultaddspace=.5em
}

\usepackage{amsmath,amssymb}
\usepackage{graphicx}
\usepackage[usenames,dvipsnames]{xcolor}
\usepackage{amsthm}
\usepackage{booktabs}
\usepackage{hyperref}
\usepackage[printwatermark]{xwatermark}
\bibpunct{[}{]}{,}{n}{}{}
\hypersetup{colorlinks=true,citecolor=Red,linkcolor=Blue,urlcolor=Black}
\usepackage[export]{adjustbox}

\begin{document}

\title {Signatures of pairing in the magnetic excitation spectrum of strongly correlated ladders}

\author{A. Nocera}
\affiliation{Department of Physics and Astronomy, The University of Tennessee, Knoxville, 
Tennessee 37996, USA}
\affiliation{Materials Science and Technology Division, Oak Ridge National Laboratory, Oak Ridge, Tennessee 37831, USA}

\author{N. D. Patel}
\affiliation{Department of Physics and Astronomy, The University of Tennessee, Knoxville, 
Tennessee 37996, USA}
\affiliation{Materials Science and Technology Division, Oak Ridge National Laboratory, Oak Ridge, Tennessee 37831, USA}

\author{E. Dagotto}
\affiliation{Department of Physics and Astronomy, The University of Tennessee, Knoxville, 
Tennessee 37996, USA}
\affiliation{Materials Science and Technology Division, Oak Ridge National Laboratory, Oak Ridge, Tennessee 37831, USA}

\author{G. Alvarez}
\affiliation{Computational Science and Engineering Division and Center for Nanophase Materials Sciences, Oak Ridge National Laboratory, Oak Ridge, Tennessee 37831, USA}

\begin{abstract}
Magnetic interactions are widely believed to play 
a crucial role in the microscopic mechanism leading 
to high critical temperature superconductivity.
It is therefore important to study the signatures of 
pairing in the magnetic excitation spectrum of simple models known to show unconventional 
superconducting tendencies. Using the Density Matrix Renormalization 
Group technique, we calculate the dynamical spin structure factor $S({\bf k},\omega)$
of a generalized $t-U-J$ Hubbard model 
away from half-filling in a two-leg ladder geometry. 
The addition of $J$ enhances pairing tendencies.
We analyze quantitatively the signatures of pairing in the magnetic 
excitation spectra.
We found that the superconducting pair-correlation strength, that can be estimated
independently from ground state properties, is closely correlated with the integrated
low-energy magnetic spectral weight in the vicinity of $(\pi,\pi)$. In this wavevector
region, robust spin incommensurate features develop with increasing doping.
The branch of the spectrum with
rung direction wavevector $k_{rung}=0$ does not change substantially with doping where 
pairing dominates, and thus plays a minor role.
We discuss the implications of  our results for neutron scattering experiments,
where the spin excitation dynamics of hole-doped quasi-one 
dimensional magnetic materials can be measured, and also address implications
for recent resonant inelastic X-ray scattering experiments.
\end{abstract}

\maketitle

\section{Introduction}

Magnetism is believed to play a key role in the pairing mechanism 
leading to high critical temperature superconductivity~\cite{scalapino12,dagotto94,johnston10,chubukov12,dai12}. 
In several materials, the neutron scattering technique is a powerful tool 
to study magnetic excitations because it can help to identify
both their energy and momentum dependence over the 
entire Brillouin zone. 
For a wide range of high critical temperature superconductors, including 
cuprates~\cite{bourges2005resonant}
and pnictides~\cite{christianson2008unconventional,lumsden2009two,PhysRevLett.103.067008}, 
an interesting feature of the magnetic excitation spectrum  
in the superconducting phase is the presence of a resonance peak
at a particular wave-vector transfer. In addition, in some cuprates spin incommensurate peaks
that develop upon doping have been associated with the presence of stripes~\cite{stripes}.
In general, a key challenge in the field of high-$T_c$ superconductivity 
is to distinguish between universal and nonuniversal properties.
In the cuprates, understanding the relationship between the magnetic resonance peak 
and superconductivity is made difficult by the occurrence 
of charge stripes and the pseudogap phase.
Recent experiments have mapped out the spin excitations in 
various cuprate families over a large range of energies~\cite{tranquada2004quantum,hinkov2004two,hayden2004structure,stock2005incommensurate,pailhes2004resonant},
 pointing toward a universal spin excitation spectrum~\cite{tranquada2006universal}
characterized by an ``hour-glass'' 
shape with a high-intensity peak at wave vector $(\pi,\pi)$, and both downward
and upward dispersing branches of excitations resembling spin incommensurate features.

In early studies of iron-based superconductors a simple picture 
dominated~\cite{christianson2008unconventional,lumsden2009two,PhysRevB.79.134520,RevModPhys.87.855}
mostly due to their more itinerant nature. 
The continuum of magnetic excitations is gapped in the superconducting 
state, and the magnetic resonance occurs at an
energy below the gap, because of the
unconventional symmetry of the superconducting order parameter, 
and the residual interaction between the quasiparticles that shifts the
pole in the total susceptibility to lower energies \cite{PhysRevB.79.134520}.
Even if in most iron-based 
superconductors the magnetic resonance has been observed at commensurate wave vectors, 
recent studies have shown that in doped compounds the resonance could be found
at incommensurate wave-vector transfers~\cite{argyriou2010incommensurate}.
However, note that the most recent developments in the field or iron-based superconductors
have revisited the weak coupling approximation and Fermi surface
nesting rationale~\cite{dai12}. 
For example there are superconducting compounds that only have electron pockets
at the Fermi level~\cite{liu12}. In fact, evidence is accumulating that pnictides and chalcogenides are in the
difficult intermediate coupling regime where neither a fully itinerant nor a fully localized picture is valid.

To better guide experiments,
it is therefore of considerable importance to investigate theoretically the magnetic 
excitations spectra of model Hamiltonians that present 
unconventional superconducting tendencies in their ground states. However,
this task is technically formidable. In layered geometries, there are no reliable
computational techniques to address the ground state properties of the
system doped away from half-filling
at the low temperatures characteristic of superconductivity. For example,
Quantum Monte Carlo techniques suffer from sign problems.
In addition, the study of dynamical magnetic spectral functions, such as
$S({\bf k},\omega)$, are also challenging due to the limitations of Maximum
Entropy procedures.

For all these reasons, it is imperative to find a simple example where computational
techniques allow for the simultaneous and accurate calculation of both
ground state pairing properties as well as dynamical spectral functions.
In this publication, we provide the first steps in this direction by
carefully analyzing the Hubbard model defined on a two-leg ladder, supplemented by a superexchange $J$
to boost further the pairing tendencies, using the Density Matrix Renormalization
Group (DMRG) technique~\cite{white92}, 
both addressing the ground state as well as its dynamical
properties. Our main challenge is: can we identify
features in the dynamical spin-structure-factor 
that appear proportional to the pairing strength?
In this work, we will show that \emph{in the case of two-leg ladders 
ground state pairing properties
are correlated with the low-energy spin excitations spectral weight of the system 
close to the magnetic wave-vector transfer $(\pi,\pi)$}.

In the 90's, model Hamiltonian studies of copper-oxide two-leg ladders 
were fruitful in elucidating 
several physical properties of their two dimensional 
counterparts~\cite{dagotto1995surprises,dagotto1999experiments}.
Experimentally, the intrinsically doped
Sr$_{2}$Ca$_{12}$Cu$_{24}$O$_{41+\delta}$ two-leg ladder material was found to become 
superconducting under pressure~\cite{uehara1996superconductivity}, 
establishing a strong link between two-leg ladders and layer-based cuprates.
Indeed, in striking similarity with the underdoped 
two-dimensional cuprates, the doped $t-J$ and Hubbard ladders show superconducting tendencies, 
as described in~\cite{dagotto1995surprises,dagotto1999experiments,PhysRevB.45.5744,tsunetsugu1994pairing,maier2008dynamics,poilblanc2003superconducting,hayward1995evidence}. 
Recently, the authors of~\cite{PhysRevB.92.195139} have revisited the crucial 
question of which is the dominating instability 
in doped Hubbard ladders employing state-of-the-art computational procedures, 
concluding in favor of pairing in the limit of small doping.

The study of the dynamical magnetic properties of superconducting ladders have 
received less attention. 
In~\cite{poilblanc2004resonant}, the dynamical 
spin structure factor of doped two-leg $t-J$ ladders were studied, concluding
that a hole pair-magnon bound-state 
evolves into a magnetic resonant excitation at finite hole doping. 
That study was performed with Lanczos exact diagonalization on 
$L=12\times 2$ clusters. Ring exchange terms were proven to be 
important to understand the spin dynamics of insulating cuprate
materials~\cite{matsuda2000magnetic,nunner2002cyclic,PhysRevLett.86.5377}.
With this motivation, in~\cite{roux2005doped}  
the dynamical spin spectrum of doped $t-J$ ladders with 
ring exchange interactions were studied as well.
Recently, an analytical low-energy 
effective field theory description of the doped Hubbard two-leg ladder model
also reported~\cite{essler2007dynamical} an incommensurate coherent mode near $(\pi,\pi)$.

The present work aims to fill the gaps in the above mentioned literature by investigating 
in detail the dynamical spin spectrum of a generalized $t-U-J$ model using the DMRG method.
In this regard, it was argued that in a pure Hubbard model 
the exchange correlation strength is constrained by the local 
Coulomb repulsion to be proportional to $J \sim t^2/U$, and pairing 
tendencies are difficult to observe due to the competition with 
other phases (CDW or stripe-like phases). If the pairing correlation strength is 
linked to the effective exchange interaction strength, 
this would tend to zero in the limit of infinite Coulomb repulsion.
For this reason, it was proposed~\cite{daul2000pairing,PhysRevB.63.100506,
PhysRevLett.90.207002,PhysRevB.79.014524} that a more realistic model 
of the cuprates is given by a generalized $t-U-J$ model, because this model
allows for an exchange magnetic interaction $J$ that is independent 
of $U$. 

In a two-leg ladder geometry---the main focus of this paper--- previous efforts
found~\cite{daul2000pairing} that pairing tendencies are enhanced as
 ``extra'' superexchange interactions $J$ are added to the Hubbard model.
While the pairing tendencies in the ground state after introducing $J$ 
have been studied~\cite{arrachea2005u,daul2000pairing,abram2013d},
the computation of the magnetic excitation spectrum of the $t-U-J$ model has 
not been reported until now.

The main aim of this paper is the following. We wish to analyze whether the magnetic spectrum
of two-leg ladders displays features that are correlated with 
the pairing tendencies known to be present in the ground state.
In other words, we wish to establish a correspondence between pairing properties, directly measured
in the ground state, with properties of the magnetic spectrum that can be measured via neutron
scattering experiments. Our main conclusion is that varying hole doping, the low-energy 
integral of the magnetic spectral weight in the vicinity of $(\pi,\pi)$ correlates qualitatively 
with the pairing correlation strength deduced from pair-pair correlations in the ground state. We also observed that
the portion of the magnetic spectrum related with wavevector $0$ along the rung direction, as opposed
to $\pi$, does not
seem related to pairing.

Our results have implications not only for neutron scattering but also
for resonant inelastic X-ray scattering (RIXS) experiments. 
Indeed, RIXS has recently emerged as a complementary tool 
to neutron scattering to study the magnetic excitations 
of strongly correlated materials~\cite{re:Ament2011}. 
In particular, recent RIXS investigations of La$_{2-x}$Sr$_{x}$CuO$_{4}$ (LSCO) have reported the persistence 
of high energy magnetic excitations from the underdoped up to the highly overdoped regime where superconductivity 
disappears~\cite{dean2012spin,dean2013persistence,re:Wakimoto2015},
raising questions about the role of these excitations  
in the pairing mechanism for cuprates. In contrast, neutron 
scattering experiments have shown that low energy 
magnetic excitations around the antiferromagnetic zone center 
``disappear'' with sufficient hole doping~\cite{re:Wakimoto2007}.
These contrasting results have been reconciled theoretically in~\cite{jia2014persistent},
confirming the persistence of high energy magnetic excitations along the 
antiferromagnetic zone boundary 
while pointing to the important role of magnetic excitations around the antiferromagnetic zone center 
$(\pi,\pi)$ in the pairing mechanism. Recently, another theoretical 
investigation~\cite{huang2017decrease} of the two-dimensional Hubbard model employing
Quantum Monte Carlo and Maximum Entropy techniques
varying doping has confirmed that high energy magnetic excitations are marginal 
to the pairing mechanism, while the main reason for the reduction of the pairing 
strength (and of the superconducting 
transition temperature $T_c$) is related with the redistribution of spectral 
weight at wave-vector momentum transfers not accessible to RIXS experiments. 
In~\cite{re:Meyers2017}, RIXS and neutron scattering 
measurements of the same LSCO sample demonstrated that the two techniques can probe magnetic excitations in 
complementary regions of the Brillouin zone, showing that the contrast between the 
results obtained with the two approaches could be solved also experimentally in the future.

This work is organized as follows. Section~II introduces the $t-U-J$ model, 
and briefly reviews its known limiting cases: the Hubbard and $t-J$ models. 
Section~III contains the main results.
In sections~III.A and III.B we present the ground state properties and the magnetic excitation 
spectrum of a $t-U-J$ ladder at 
fixed realistic hole doping and Coulomb repulsion, changing the magnetic exchange interaction 
$J$. Section~III.C explores the properties of the magnetic excitations at the same doping 
but now as a function of local Coulomb repulsion $U$, at $J=0.0$.
Section~III.D contains our main results, namely
the dependence of the magnetic excitation 
spectrum as a function of the hole doping, and its correlation with ground state pairing properties.
In Section~IV we provide our conclusions.

\section{Generalized Hubbard Model}

The Hamiltonian of the generalized $t-U-J$ model defined on a two-leg ladder geometry is
\begin{align}\label{eq:HtJUladder}
H&=\Big(-t_{x} \sum\limits_{\substack{\langle i,j\rangle\\ \sigma,\gamma=0,1}}
c^\dagger_{i,\gamma,\sigma}
c_{j,\gamma,\sigma} - t_y \sum\limits_{i,\sigma} 
c^{\dag}_{i,0,\sigma}c_{i,1,\sigma}\Big)+ \text{h.c.} \nonumber\\
&+ U\sum\limits_{i,\gamma=0,1} n_{i,\gamma,\uparrow}n_{i,\gamma,\downarrow}+
J_x \sum\limits_{i,\gamma=0,1} \Big(\vec{S}_{i,\gamma}\cdot\vec{S}_{i+1,\gamma}+\nonumber\\
&-\frac{1}{4} n_{i,\gamma} n_{i+1,\gamma}\Big) + J_y \Big(\sum\limits_{i,\sigma} 
\vec{S}_{i,0}\cdot\vec{S}_{i,1}-\frac{1}{4} n_{i,0} n_{i,1}\Big),
\end{align}
where $c^\dagger_{i,\gamma,\sigma}$ ($c_{i,\gamma,\sigma}$) creates 
(destroys) an electron 
at leg $\gamma=0,1$ on site $i=0,L/2-1$ and spin $\sigma=\uparrow,\downarrow$. 
$n_{i,\gamma}$ and $\vec{S}_{i,\gamma}$ represent the electronic occupation operator 
(summed over spins) and spin operators on site $i$ and leg $\gamma$. 
Following standard notation, $t_x$ and $t_y$ represent the hopping parameters 
in the $x$ (along the leg) and $y$
(along the rung) direction of the ladder. For simplicity, in most figures the $y$ direction
wavevector will be explicitly indicated as $k_{rung}$ while the $x$ direction wavector will be denoted as $k$.
$J_x$ and $J_y$ are the exchange interactions along the leg and 
rung directions, respectively. $U$ is the local Hubbard on-site Coulomb repulsion strength.
We consider $t_x=t_y=1$ as unit of energy, and $J_x=J_y=J$ for the
exchange interaction. The model above reduces to the standard Hubbard 
model for $J=0$, and to the standard $t-J$ model
for $U\rightarrow\infty$, as double electronic occupation 
is forbidden in this limit.

Let us recall the basic properties of the standard 
Hubbard ladder,  recovered from Eq.~(\ref{eq:HtJUladder}) in the $J=0$ limit. 
In the symmetric $t_x=t_y$ and non-interacting $U=0$ case, 
both bonding ($+$) and antibonding bands ($-$), 
$\epsilon_{\pm} = -2(\cos(k_x)\mp 1)$, are filled up by 
electrons if the electronic density $n=N_{el}/L$ is larger than quarter-filling, 
$n=0.5$, where $N_{el}$ is the total number of electrons and $L$ 
the total number of sites, $L/2$ for each leg. 
Only the case of $n\geq 0.50$ will be considered in this paper. 
When $n\geq 0.50$, there are then four Fermi points: $\pm k_{F_+}$ for
the bonding and $\pm k_{F_-}$ for the anti-bonding bands. 
At generic filling $n$, $k_{F_+} + k_{F_-} = \pi n$.
At half-filling and $U>0$ the Hubbard ladder has both a charge and spin gap.
Away from half-filling, the charge gap disappears, while the spin gap decreases 
remaining finite up to large finite 
doping~\cite{re:Noack1995,re:Noack1996,re:Endres1996,jeckelmann1998comparison}. 
Moreover, the system presents power law $d_{x^2-y^2}-$like pair-field 
correlations~\cite{noack1997enhanced,PhysRevB.92.195139}.

The $t-J$ model on ladders has been thoroughly studied in the cuprates 
literature~\cite{PhysRevB.45.5744,dagotto1995surprises,dagotto1999experiments}.
In the undoped limit, it has been well established that the $t-J$ model 
has a spin gap due to the particular ladder geometry which favors the spin singlet
formation along rungs, and the physics can be well described in terms of the 
Heisenberg ladder model. Upon doping,
superconducting tendencies develop~\cite{PhysRevB.45.5744,dagotto1995surprises,tsunetsugu1994pairing,maier2008dynamics,poilblanc2003superconducting,hayward1995evidence}. 
The physics 
of $t-J$ (and Hubbard) two-leg ladders has been studied with many techniques ranging 
from Exact Diagonalization to DMRG to bosonization~\cite{re:White2002}.
Away from half-filling,
the spin gap and superconducting binding energy of hole pairs was studied in~\cite{re:Riera1999}
showing that
they can be maximized by tuning the anisotropic ratios to $t_y/t_x \simeq 1.25$ and 
$J_y/J_x \simeq 1.56$.
In general, and important for the goals of our present publication, 
in~\cite{re:Scalapino1998} it was explained that
neutron scattering data could provide important evidence for 
the pairing mechanism based on the exchange interaction $J$. For this reason we aim
to study in parallel the pairing ground state properties of ladders introducing doping, 
as well as the inelastic neutron scattering spectrum $S({\bf k},\omega)$ under similar
circumstances and analyze whether correlations among them can be established.

The DMRG correction-vector method has been used throughout 
this paper~\cite{re:Kuhner1999}. 
Within the correction vector approach, we use the Krylov 
decomposition~\cite{nocera2016spectral}
instead of the conjugate gradient. An application of the method to Heisenberg and Hubbard 
ladders at half-filling can be found in~\cite{nocera2016magnetic}.
In this work, a $L=48\times2$ ladder has been 
simulated, using $m=1000$ DMRG states with a truncation error kept below $10^{-5}$. 
The spectral broadening in the correction-vector approach 
has been considered fixed at $\eta=0.08t$.
The DMRG implementation used throughout this paper has been 
discussed in detail in~\cite{nocera2016magnetic}; 
technical details are in the Supplemental Material~\cite{re:supplemental}.

\section{Results}

\subsection{Enhancement of pairing tendencies by $J$}

We begin by studying the ground state properties of a 
$L=48\times 2$ ladder at a fixed doping $n=0.875=84/96$.
This filling was chosen because it has been widely used
to study pairing tendencies on ladders, such as in~\cite{daul2000pairing}, 
and because the spectrum features to be described below are sharp
and clearly visible. Thus $n=0.875$ is an ideal 
doping for a preliminary understanding of the dynamical structure factor.

\begin{figure}[h]
\includegraphics[width=8.5cm]{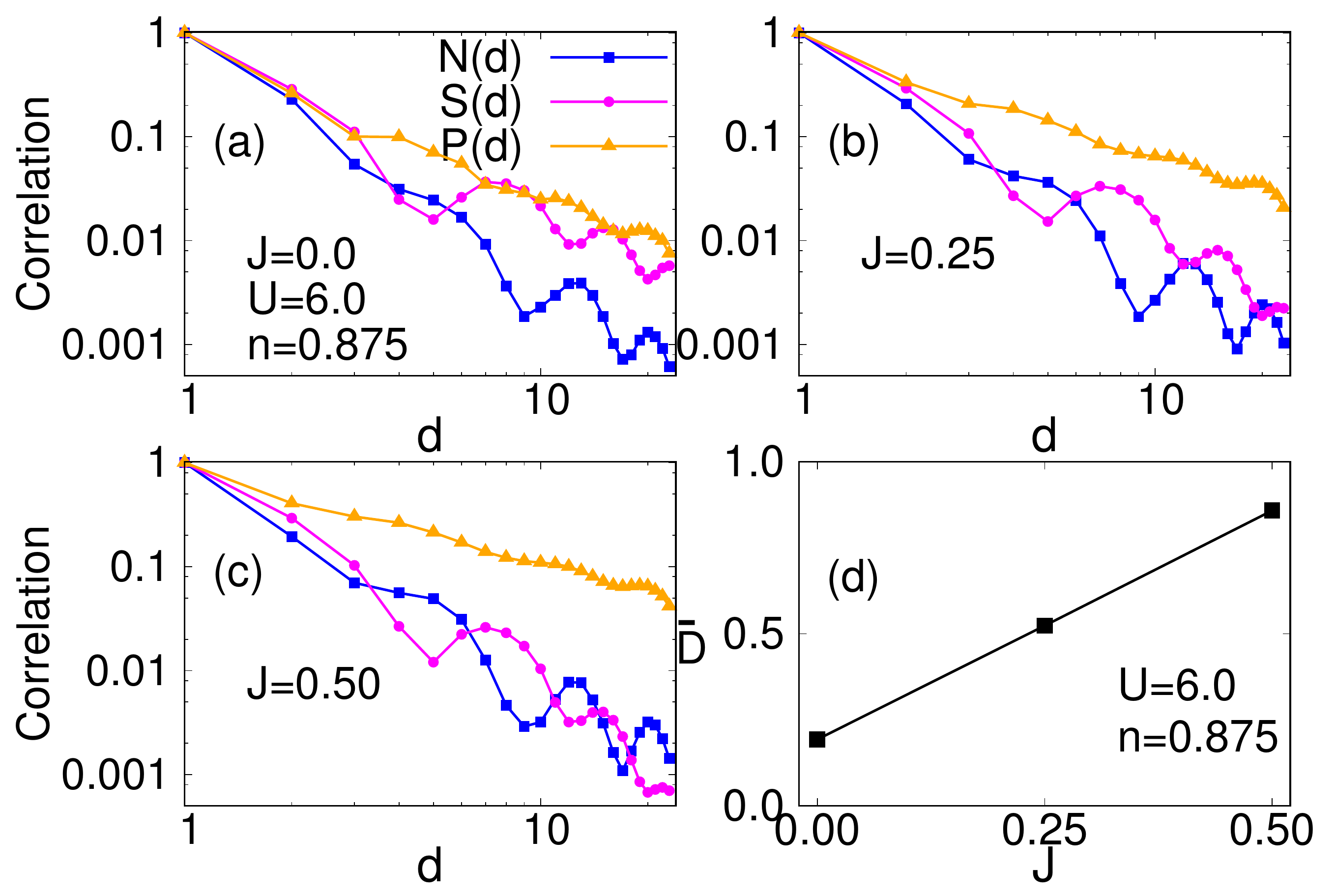}
\caption{(Color online) Panels (a-b-c): Spin, density 
and rung-singlet pair correlation functions 
versus distance along the ladder leg for different values 
of the magnetic exchange interaction $J$. Note that in these
panels the results are normalized to those at distance 1. 
Panels (d): Pairing correlation strength (see text) 
as a function of magnetic 
exchange interaction $J$. A $L=48\times2$ ladder has been 
simulated, with electronic filling $n=0.875=84/96$. 
The number of DMRG states kept is $m=1000$.} \label{fig:1}
\end{figure}

We have checked that our results are in agreement with 
an early study of the same model~\cite{daul2000pairing}
for a shorter system size, $L=32\times 2$.
Figure \ref{fig:1}(a-b-c) shows the \emph{averaged} 
rung-singlet pair correlation function $P(d)$, rung spin $S(d)$, 
and rung density correlation $N(d)$ calculated as a function of the distance $d$ along the leg of the 
ladder for different values of the magnetic exchange interaction $J$, 
fixing the Coulomb repulsion to an intermediate value $U=6.0$. 
The \emph{averaged} correlation function for a generic operator $\hat{O}$ 
is defined as 
\begin{equation}
O(d) = \frac{1}{L-d}\sum_{j=1}^{L-d}\langle\hat{O}^{\dagger}_{j}\hat{O}_{j+d}\rangle.
\end{equation}  
We have a rung-singlet pair correlation $P(d)$ when $\hat{O}~=~\Delta_{i}$, with
the operator $\Delta^{\dagger}_{i}$ defined as 
\begin{equation}
\Delta^{\dagger}_{i} 
= \frac{1}{\sqrt{2}}\Big(c^{\dagger}_{i,0,\uparrow}c^{\dagger}_{i,1,\downarrow}
-c^{\dagger}_{i,0,\downarrow}c^{\dagger}_{i,1,\uparrow}\Big).
\end{equation}
For the rung spin correlations we have used
$\hat{O}_i~=~\sum_{\gamma=0,1}\vec{S}_{i,\gamma}$, 
while for the rung density correlation $N(d)$ we employed
$\hat{O}_i~=~\sum_{\gamma=0,1}n_{i,\gamma}$.

Our results in Fig.~\ref{fig:1}(a-b-c) 
confirm that the ``extra'' exchange interaction 
$J$ increases the strength of the pairing, and induces 
a slower decay of the rung-singlet pair correlation function as compared with the case $J=0$.
Moreover, Figure~\ref{fig:1} shows also that the rung 
density and rung spin correlation functions have a faster decay 
than the pair correlations. 
We can therefore conclude that the increase of the magnetic 
exchange interaction $J$ \emph{increases pairing to a point 
where it dominates}.  
This is also shown in Fig.~\ref{fig:1}(d) where the 
pairing correlation strength is estimated by evaluating the quantity 
$\bar{D}=\sum_{i=6}^{12}P(i)$ (note that $6$ and $12$ are $arbitrary$ lower and upper bounds in the sum
but we have observed that qualitatively the results are similar modifying those limits:
choosing $6$, as opposed to e.g. 1, reduces artificial short-distance effects while $12$,
as opposed to e.g. 24, reduces edge effects).
In fact $\bar{D}$  increases approximately linearly as a function 
of $J$. Our results are also in agreement with~\cite{PhysRevB.92.195139},
where a careful size scaling analysis of the correlation function 
was performed on a doped Hubbard ladder, concluding that 
superconducting correlations are dominant in the regime 
that we also investigate in this work.

\subsection{Magnetic excitations at fixed hole doping changing $J$}

\begin{figure}[h]
\includegraphics[width=8.5cm]{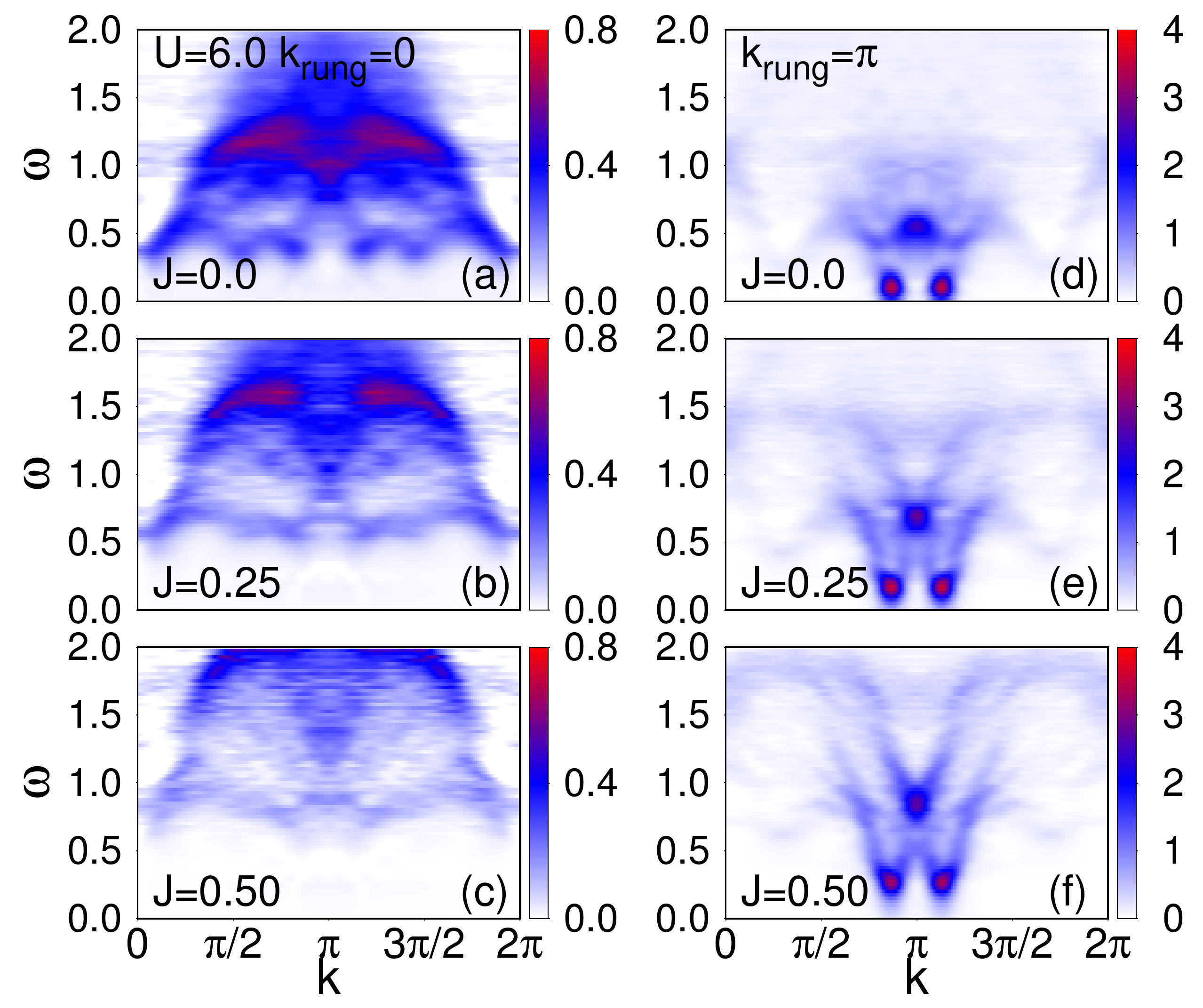}
\caption{(Color online) Panels (a-b-c): $k_{rung}=0$ 
component of the magnetic excitation spectrum for a $t-J-U$ 
ladder using $L=48\times2$ sites and the DMRG technique
at $U=6$ and filling $n=0.875=84/96$. 
Panel (a) corresponds to $J=0.0$, (b) to $J=0.25$, and (c) to $J=0.50$. 
Panels (d-e-f): $k_{rung}=\pi$ 
component of the magnetic spectrum for the same parameters used above. 
The number of DMRG states kept is $m=1000$.
} \label{fig:2}
\end{figure}

Figure~\ref{fig:2} shows the dynamical spin structure factor 
of the ladder changing the value of the exchange interaction 
$J$ for the same parameters investigated above.

Figures~\ref{fig:2}(a) and (d) show the magnetic excitation 
spectrum for a doped Hubbard ladder at $J=0.0$. 
In the $k_{rung}=0$ component (a), an umbrella-like 
shape manifold of excitations appears 
above a robust gap, for the
value of doping chosen. 
As the exchange magnetic interaction $J$ increases, 
one can observe that the region of depleted weight
 at $k_{rung}=0$ also  increases and
the entire spectrum is pushed up in energy, but the 
shape of the dispersion of the low-energy excitation band 
does not change qualitatively with $J$. However, the amplitude of the low energy 
sinusoidal oscillations in the magnetic dispersion are damped by the extra magnetic 
exchange interactions.  

In the $k_{rung}=\pi$ 
component of the magnetic spectrum~Fig.~\ref{fig:2}(d), 
most of the spectral weight is concentrated at the incommensurate 
wave-vector $k_x\simeq\pi n$ (with the companion peak 
at $2\pi-\pi n$; $x$ is the leg direction) as it can also be
observed in the static structure factor (not shown). 
These sharp peaks do not seem to be separated in energy from the rest 
of the spectrum at higher energy, at least within the resolution of our study. 
Indeed, our data does not contain features that could be associated with a
bound state magnetic excitation distinctly separated below a
continuum of excitations at this hole-doping density,
and we have also verified 
this observation for a shorter system size, $L=32\times2$. Namely,
we cannot distinguish a clear ``resonance'' feature in the spin excitations
even though ground state measurements indicate that pairing tendencies 
are dominant. However, although the subtle issue of the existence of a resonance 
merits further elaboration, our focus in the rest of the manuscript 
is quite different, as explained below.
Figure~\ref{fig:2}(d) suggests that at $J=0.0$ from each 
sharp low-energy peak a linear branch of higher energy excitations 
with smaller spectral 
weight develops, with 
reflection symmetry with respect to $k_x=\pi$. 
Another visible broad band of excitations occurs 
at high energy around $\omega/t\simeq0.5$, with a central 
peak around $k_x=\pi$.

These $k_{rung}=\pi$ spectral features described above are enhanced and
clarified when 
``extra'' exchange interactions are introduced compared to 
the pure Hubbard model obtained for $J=0.0$. With increasing $J$
the spin gap of the incommensurate spin excitations
increases, while their intensity remains approximately the same with
increasing $J$. This is interesting because $J$ increases
the pairing tendencies in the ground state. 
Moreover, we observe that by increasing $J$, spectral weight leaks to the triangle-shape area between 
the incommensurate peaks and the higher energy peak at $(\pi,\pi)$ which occurs around $\omega\simeq0.9$
for $J=0.5$. As we will explain in section III.D, we will use 
the low energy spectral weight around $(\pi,\pi)$ as a measure of the pairing correlation strength in the system 
and we will find that it increases with $J$.
As the exchange magnetic interactions are increased, the 
$W$-shaped energy spectral feature observed in the $k_{rung}=\pi$ component 
of the magnetic excitation spectrum at $\omega/t\simeq0.5$ for $J=0.0$
is pushed at higher energies, with two long $V$-shape spectral bands developing,
starting from the incommensurate low energy peaks. 
These two spectral features intersect at $(\pi,\pi)$ at higher 
energy transfer which increases with $J$. 
We can summarize the results of this section by stating that 
the ``extra'' exchange interaction $J$ 
mainly shifts the magnetic spectrum to higher energies, maintaining the 
dispersive features of the magnetic excitations qualitatively unchanged.

\subsection{Magnetic excitations at fixed hole doping changing $U$}


In this section, we study the magnetic excitation spectrum of the doped ladder 
($n=0.875$) at $J=0.0$ for different values of the on-site Hubbard repulsion $U$.  
By studying the ground state rung-singlet pair correlations, we have found that 
the pairing correlation strength increases with $U$ starting from $U=0.0$, 
it reaches a broad maximum in the range $U\sim 4-6$, 
and eventually decreases as $U$ is further increased. 
Analogously, we have found that also the spin 
gap has a similar behavior, reaching a maximum for $U\sim4-6$ 
as in the case of half-filling~\cite{nocera2016magnetic,daul2000pairing}.
These results agree with those reported 
in~\cite{daul2000pairing}, supporting the notion that 
Hubbard on-site repulsion reduces charge and spin fluctuations 
such that pairing dominates at intermediate $U$, while 
for very large $U$
spin fluctuations will eventually dominate over pairing.

\begin{figure}[h]
\includegraphics[width=8.5cm]{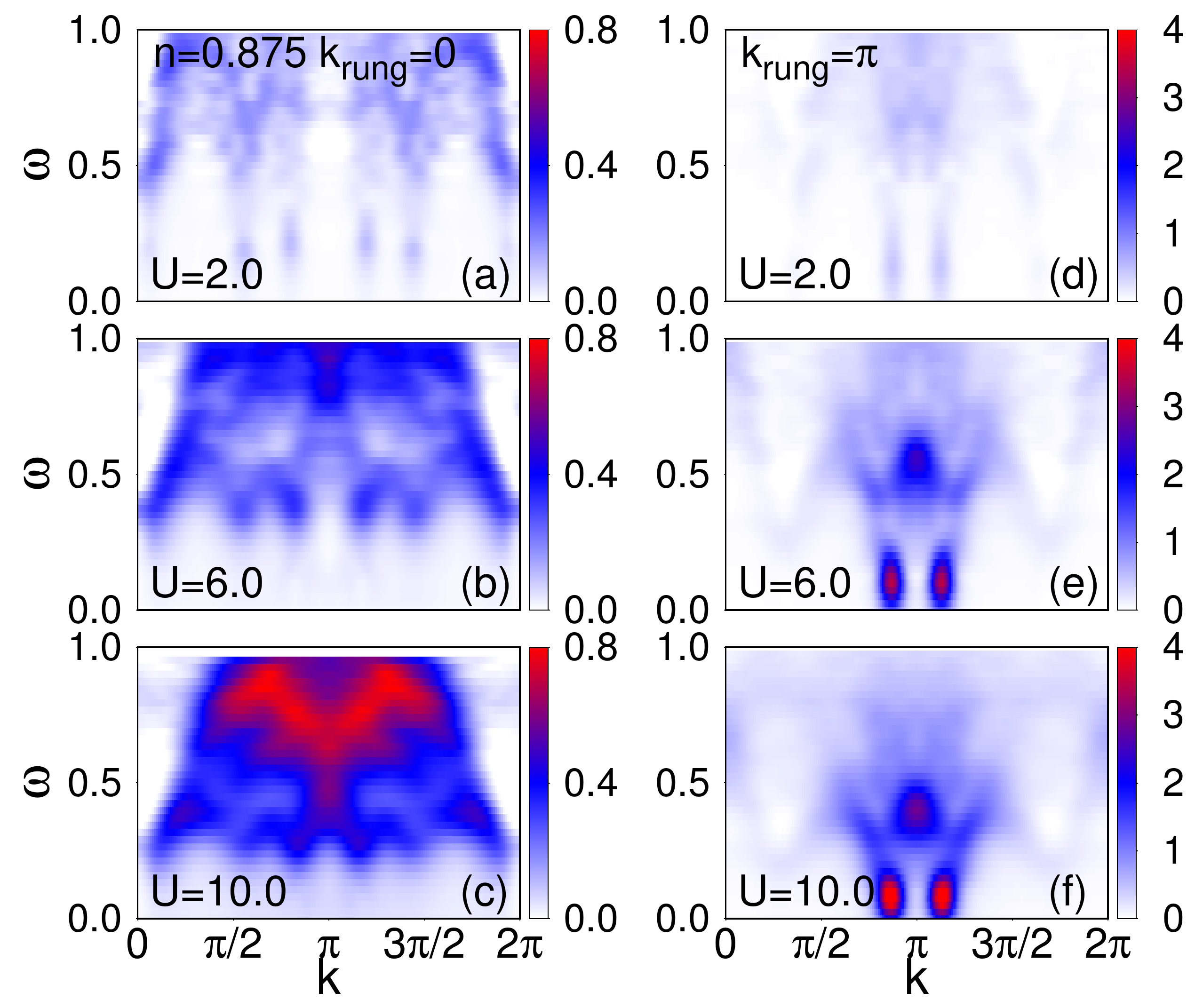}
\caption{(Color online) Panels (a-b-c): $k_{rung}=0$ 
component of the magnetic excitation spectrum obtained with DMRG for a $t-J-U$ ladder 
of $L=48\times2$ sites, at $J=0.0$ and $n=0.875$. 
Panel (a) corresponds to $U=2.0$, (b) to $U=6.0$, 
and (c) to $U=10.0$. Panels (d-e-f) are the $k_{rung}=\pi$ 
component of the magnetic spectrum, for the same parameters
as above.} \label{fig:7}
\end{figure}

Figure~\ref{fig:7} shows the dynamical spin structure factor 
of our doped ladder changing the value of $U$
at $J=0.0$. 
Panels (a,b,c) of Fig.~\ref{fig:7} 
display the $k_{rung}=0$ component of the magnetic excitation spectrum. 
At $U=2.0$ the results are still similar to the noninteracting limit.
At $U=6.0$ and beyond, there is a gap at low energy that closes
with increasing $U$, keeping approximately the overall shape of the spectrum.
There are no dominant coherent peaks. The sinusoidal oscillations of the
lower energy magnetic excitations are damped as $U$ increases. Overall,
similarly as in the previous analysis varying $J$, the $k_{rung}=0$ portion
of the spectrum contains broad features but not much coherence.

Figures~\ref{fig:7}(d-e-f) show the $k_{rung}=\pi$ 
component of the magnetic excitation spectrum 
for the same parameter values investigated above. A spectral redistribution from high energy to low energy 
is observed, as in the $k_{rung}=0$ component of the spectrum. 
In~\cite{nocera2016magnetic}, where the crossover Hubbard-to-Heisenberg behavior 
in the half-filled case was carefully studied, a similar spectral 
weight redistribution was observed. At $U=2.0$ the results resemble the non-interacting case
with very low weight in the energy range studied. As $U$ increases, sharp incommensurate
peaks develop at low energies (note the change in the intensity convention between left and right panels),
similarly as when $J$ was varied before at fixed $U=6.0$. 

We can summarize the results of this subsection by stating that 
increasing from zero the on-site Hubbard repulsion, the spectral weight 
much increases at intermediate-low energies in both branches. The $k_{rung}=0$
component remains disorganized varying $U$, but the $k_{rung}=\pi$ component
develops coherent sharp peaks at low energies that likely dominate the physics
related with the interaction between the charge and spin degrees of freedom.

\subsection{Magnetic excitations as a function of hole doping}

This section investigates the properties of the magnetic excitation spectrum for 
our $t-U-J$ ladder now as a function of hole doping, at a fixed value 
of the Coulomb interaction $U=6.0$, and attempts to correlate some of its
features with the pairing strength in the ground state.

\subsubsection{Pairing correlation strength varying $n$}

\begin{figure}
\includegraphics[width=8.5cm]{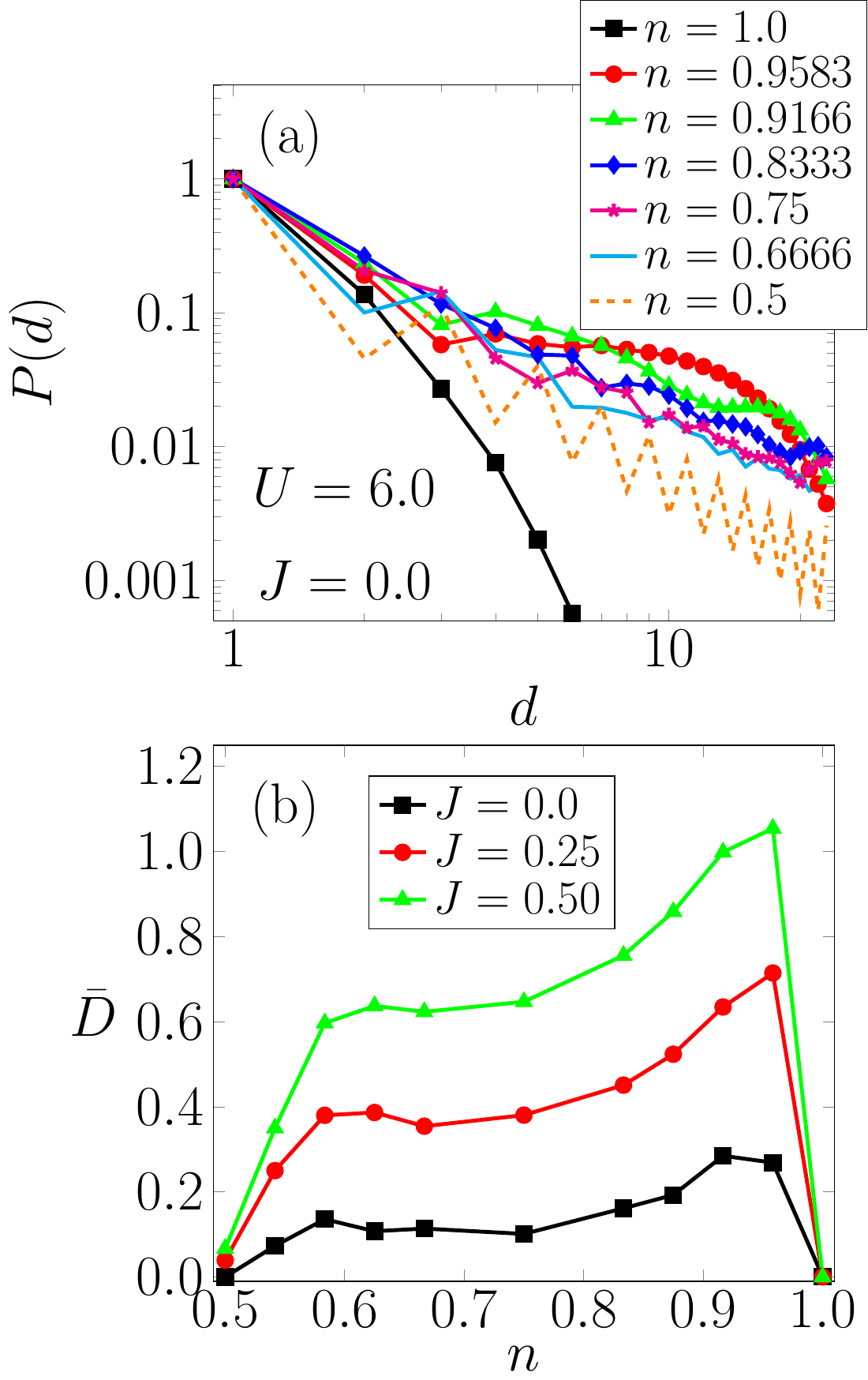}
\caption{(Color online) (a) Pair correlation functions 
corresponding to rung-singlet pairs versus distance, at several values 
of the electronic filling $n$ (see inset), and at a fixed $U=6.0$ and $J=0.0$. 
The data is normalized to distance one.
These correlations are very weak at $n=1$, rapidly develop with reducing $n$,
and then drop again at $n=0.5$. 
(b) Pairing correlation strength (defined as
$\bar{D}=\sum_{i=6}^{12}P(i)$, see text) versus electron filling $n$, at $U=6.0$ and the
values of $J$ indicated in the inset.}\label{fig:9}
\end{figure}

Figure~\ref{fig:9}(a) shows the ground state 
rung-singlet pair correlations versus distance, along the ladder leg, 
at the different values of the electronic filling indicated. Note that the results
are normalized to distance 1.
At half-filling the pair correlations decay to zero 
exponentially fast, giving a very small pairing correlation strength. 
At $J=0.0$, namely without any extra $J$ boost for hole binding, 
the results indicate that the pairing correlation strength (panel (b)) increases 
rapidly with hole doping starting from the half-filled case, develops 
a broad maximum, and then decreases reaching almost zero at $n=0.5$. The actual values
of the pairing correlation strength are  sensitive to the choices of lower and upper limits in the definition
of $\bar{D}$ but the overall shape is qualitatively stable, displaying an asymmetric superconducting dome, 
that remains approximately the same as the magnetic exchange interaction increases. Note that 
in the range from $n=0.75$ to $n=0.5833$ the pairing strength is nearly constant,
which is somewhat surprising: intuitively a monotonous decrease of $\bar{D}$ from the peak
near half-filling towards the near zero value at quarter-filling would have been more natural. 
This anomalous behavior will be addressed in future investigations.
However, our focus in the rest of the publication will be on the more robust and stable 
feature related with the clear dominant peak near half-filling and its correlation with magnetism.
  
\subsubsection{Magnetic spectrum varying $n$}

\begin{figure}[h]
\includegraphics[width=8.5cm]{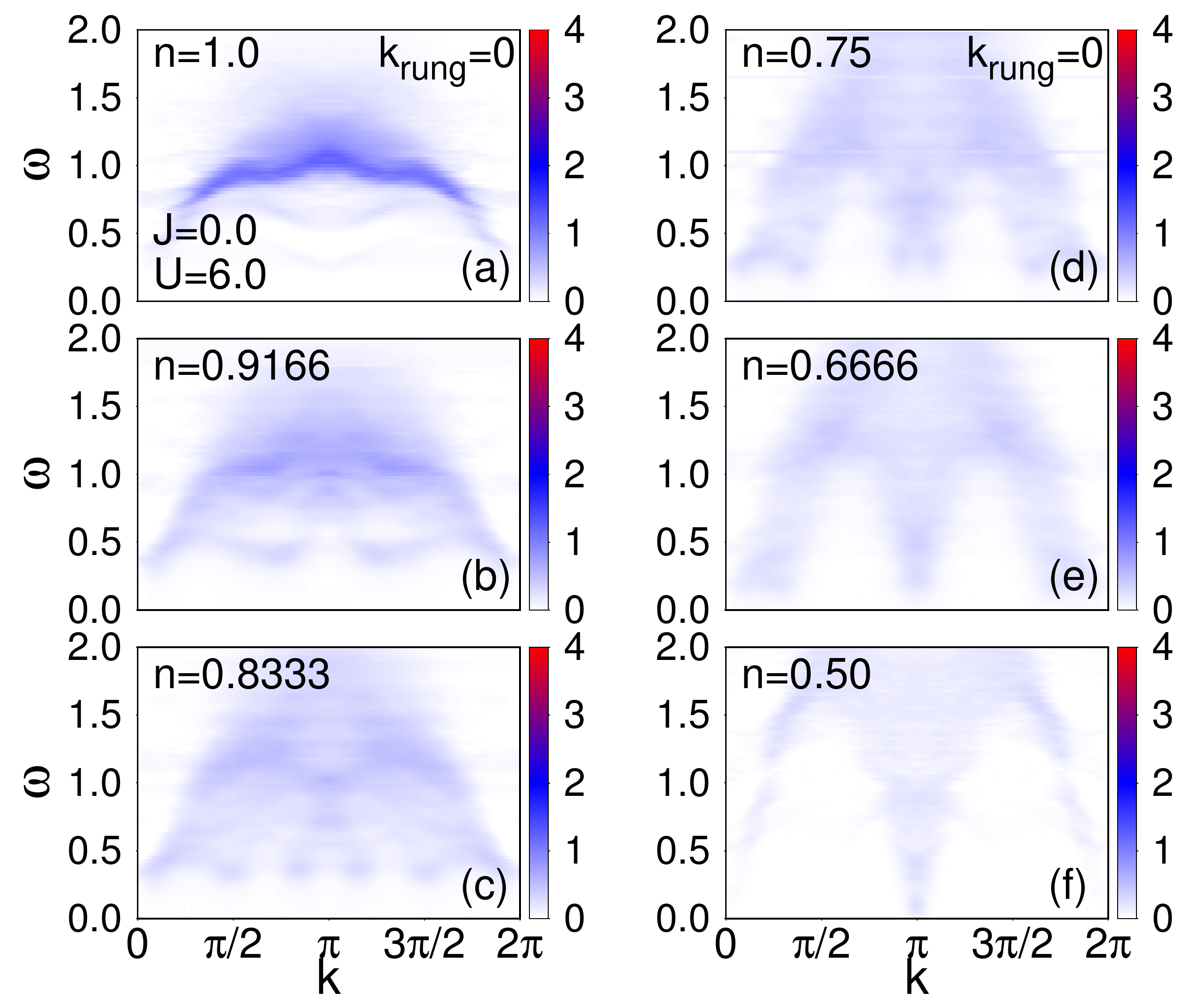}
\caption{(Color online) $k_{rung}=0$ component of the magnetic excitation spectrum 
for a $t-J-U$ ladder using $L=48\times2$ sites and the DMRG technique, at $U=6.0$ and $J=0.0$, 
and at the several electronic fillings $n$ indicated.} \label{fig:11}
\end{figure}

\begin{figure}[h]
\includegraphics[width=8.5cm]{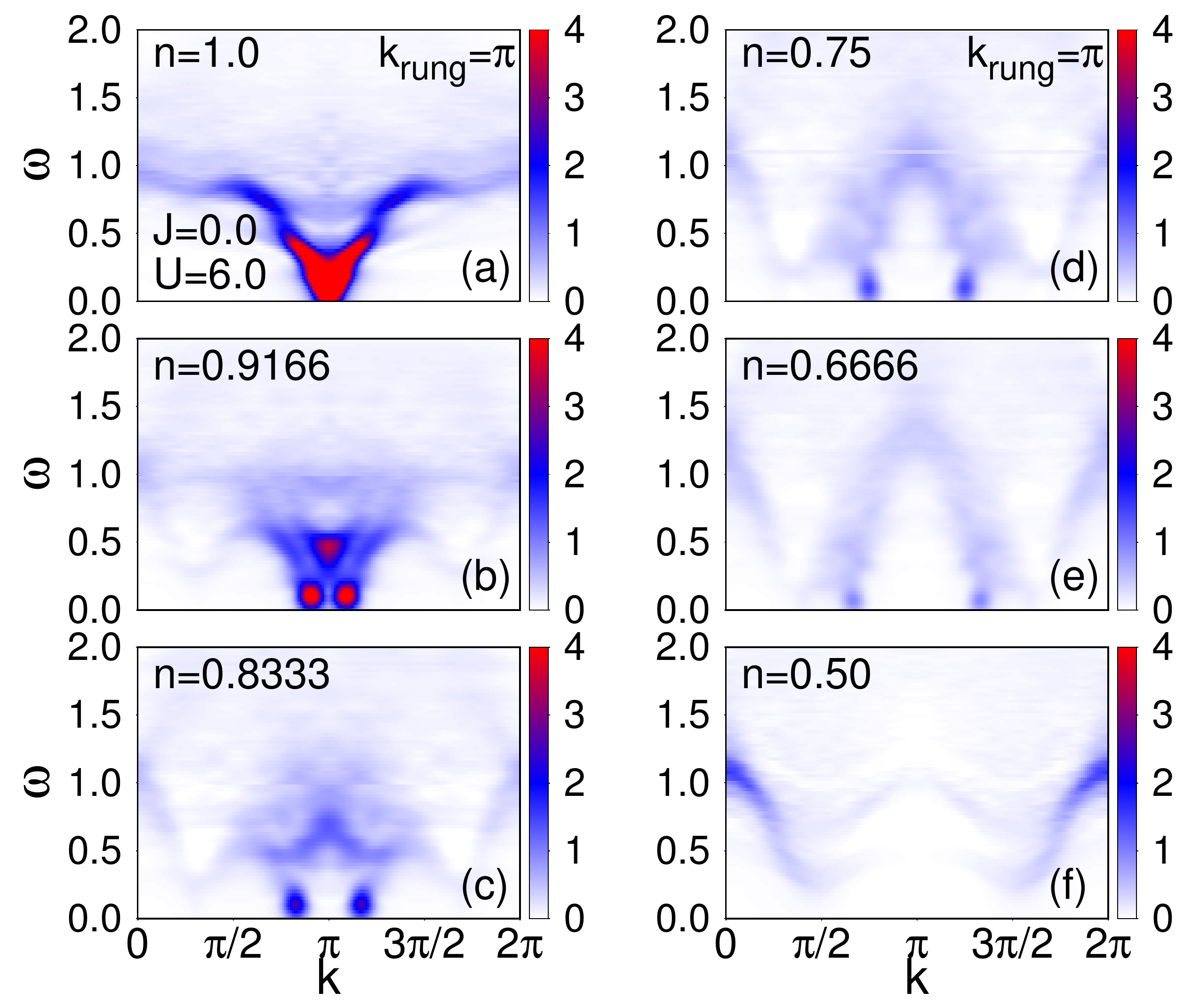}
\caption{(Color online) $k_{rung}=\pi$ 
component of the magnetic excitation spectrum 
for a $t-J-U$ ladder using $L=48\times2$ sites and the DMRG technique, at $U=6.0$ and $J=0.0$, 
and at the several electronic fillings $n$ indicated.} \label{fig:12}
\end{figure}

To study whether there are  signatures of the pairing tendencies unveiled above 
in the magnetic excitation spectrum, Fig.~\ref{fig:11} and Fig.~\ref{fig:12} 
show the $k_{rung}=0$ and $k_{rung}=\pi$ components of the 
dynamical spin structure factor at $J=0.0$, respectively, for similar values of electron 
fillings as investigated above for pairing.

Figure~\ref{fig:11} shows that by decreasing the electronic filling, 
the $k_{rung}=0$ 
component of the magnetic 
excitation spectrum maintains qualitatively its structure in the 
interval of fillings $1<n<0.8333$. 
Our results are consistent with recent studies
of the magnetic spectrum of the two-dimensional Hubbard model 
varying doping~\cite{huang2017decrease}, where it was observed that
the dispersion of the magnetic excitations along the line $(0,0)-(\pi,0)$ in 
the Brillouin zone does not change much with hole doping, while at the same time 
the pairing correlation strength was found to be reduced. Similarly, our results also indicate that the
$k_{rung}=0$ branch of the ladder spectrum does not seem correlated with pairing either.

At $n=0.75$ and lower densities, a low-intensity 
low-energy feature at the leg wavevector $k=\pi$ develops. This feature resembles results found already
in the non-interacting limit $U=0.0$ at  these densities  
indicating that this portion of the magnetic spectrum represents quasi non-interacting electrons.

Figure~\ref{fig:12} shows that the $k_{rung}=\pi$ component
contains much more of the total spectral weight of the magnetic excitations. 
In particular, we can observe at half-filling (panel (a)) the typical $V$-shape-like 
of the magnetic excitation dispersion which is characteristic of the 
Heisenberg counterpart~\cite{nocera2016magnetic} where the spectral weight is 
mainly concentrated at the scattering wave-vector $(\pi,\pi)$. In this case, the Fermi 
momentum $k_F=\pi n$ is equal to $\pi$ at half-filling ($U=6.0$ is 
strong and it is mainly accidental that such a weak coupling based on the Fermi momentum
perspective is still valid).

Upon decreasing the electronic filling, the magnetic excitation peak at $(\pi,\pi)$ splits 
in two incommensurate peaks as observed also in the previous sections at $n=0.875$, 
with separation in momentum transfer proportional 
to the electronic filling itself, $2k_F=2\pi n$. At the same time, it is evident that 
the spectral weight redistributes at smaller momentum transfers, so that overall the 
scattering region in the interval $(\pi/2,3\pi/2)$ around $(\pi,\pi)$
becomes depleted in spectral weight by hole doping. Although it can be barely noticed
in the scale used, we found that the spin gap that characterizes ladders at half-filling
is monotonically reduced with hole doping (the wavevector position that characterizes
the spin gap is that of the incommensurate features, when at finite doping). 
The fine details of how this occurs
are not important because it will be discussed below that the $integral$ of the spectral
weight near $(\pi,\pi)$ is what correlates with the ground state pairing properties.
Finally, at $n=0.5$ we observe that the 
spectral weight is mostly concentrated at $(0,\pi)$, with highly dispersive spectral 
bands appearing in the intervals $(0,\pi/2)$ and $(3\pi/2,2\pi)$.

\begin{figure}[h]
\includegraphics[width=8.5cm]{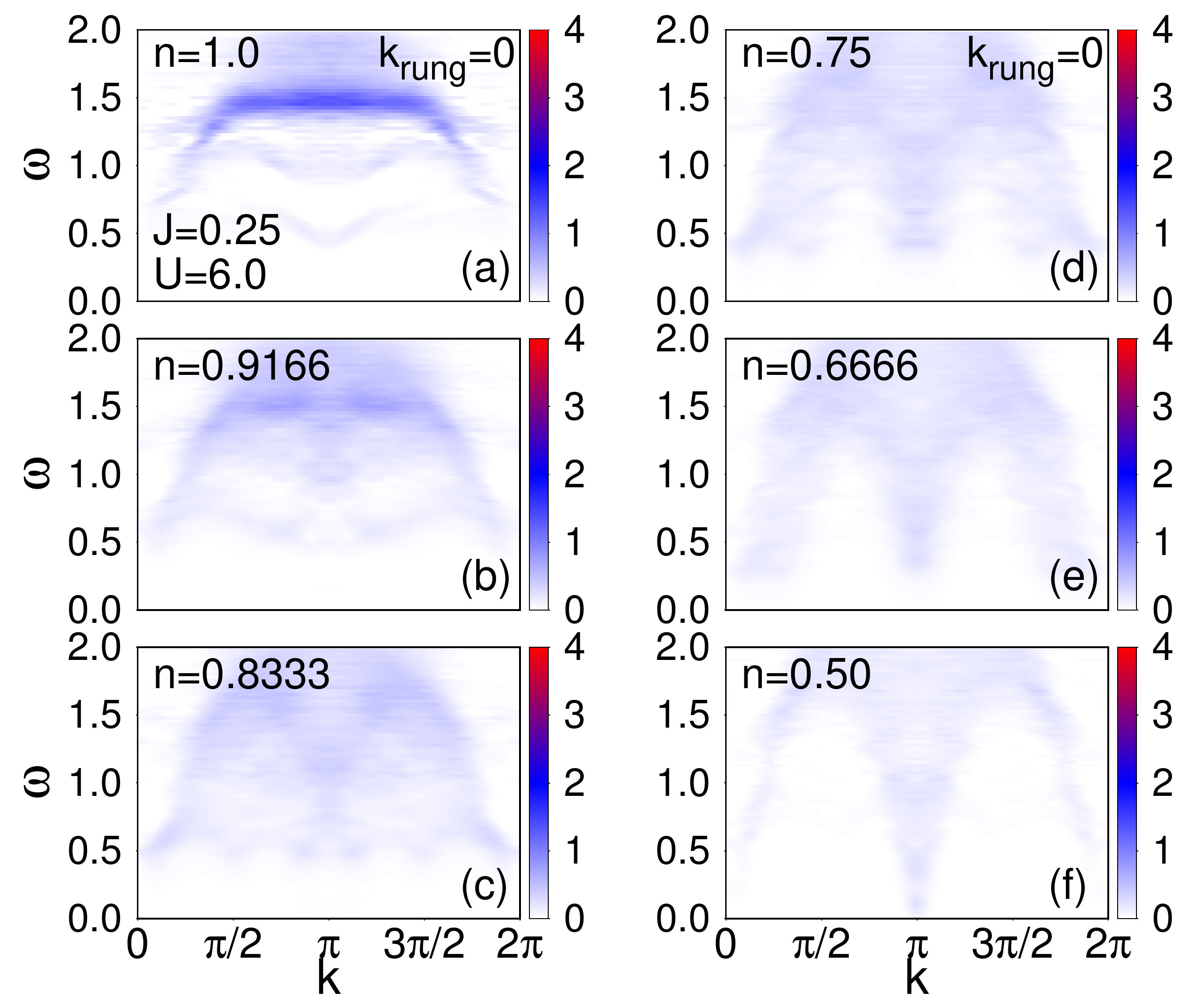}
\caption{(Color online) $k_{rung}=0$ 
component of the magnetic excitation spectrum 
for a $t-J-U$ ladder using $L=48\times2$ sites and the DMRG technique, at $U=6.0$ and $J=0.25$, 
and at the several electronic fillings $n$ indicated.} \label{fig:16}
\end{figure}

\begin{figure}[h]
\includegraphics[width=8.5cm]{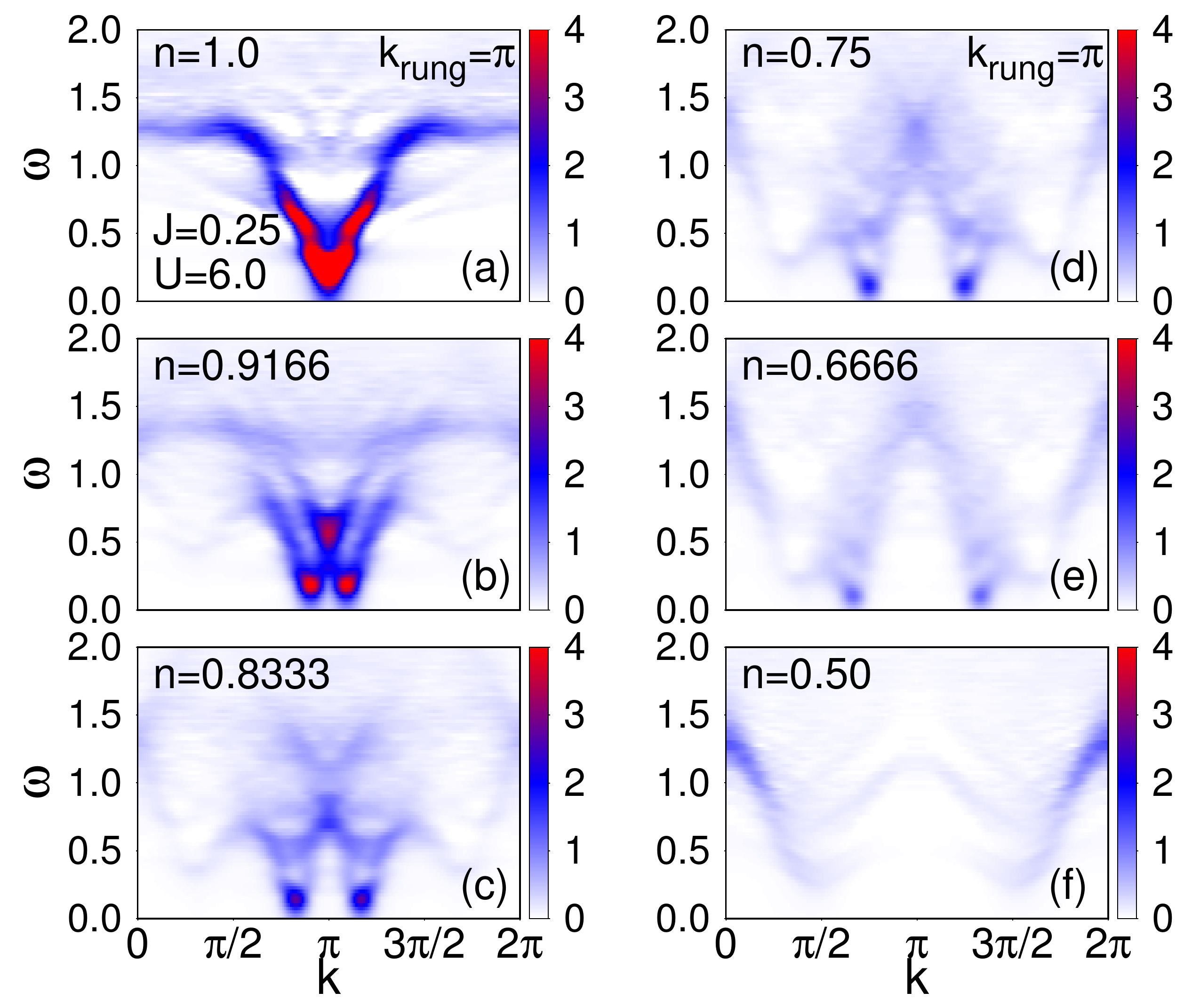}
\caption{(Color online) $k_{rung}=\pi$ 
component of the magnetic excitation spectrum 
for a $t-J-U$ ladder using $L=48\times2$ sites and the DMRG technique, at $U=6.0$ and $J=0.25$, 
and at the several electronic fillings $n$ indicated.} \label{fig:17}
\end{figure}

Considering that $J$ increases the tendencies to superconduct, it is also interesting to
analyze the magnetic spectral properties at, e.g., $J=0.25$. The results are in
Figs.~\ref{fig:16} and~\ref{fig:17}. For the $k_{rung}=0$ 
branch, the shape of the dominant
features is approximately the same as for $J=0.0$. As
expected from previous analysis, overall there is a shift to higher energies of the
spectrum.  Similar conclusions were reached for the $k_{rung}=\pi$ branch: clearly the dominant
features qualitatively are the same varying $J$. However, 
overall the intensity of the low energy
excitations increases with increasing $J$. Thus, {\it the low-energy intensity near $(\pi,\pi)$
is the property that appears the closest related to pairing}.

We can partially summarize these results by stating that  
the magnetic excitations dispersion along the line $(0,0)-(\pi,0)$ in the Brillouin zone
do not change substantially against hole doping. On the other hand,
the pairing correlation strength rapidly increases starting from half-filling $n=1.0$, 
reaching a maximum around $n\simeq0.9$ and then further decreasing by hole doping. 
At the same time, a significant spectral 
weight redistribution away from the $(\pi,\pi)$ wavevector, 
characteristic of the half-filled case, is observed as a function of hole-doping, with
the overall intensity decreasing with doping. 
It seems that pairing and the overall weight near $(\pi,\pi)$ are correlated.

\begin{figure}
\includegraphics[width=8.5cm]{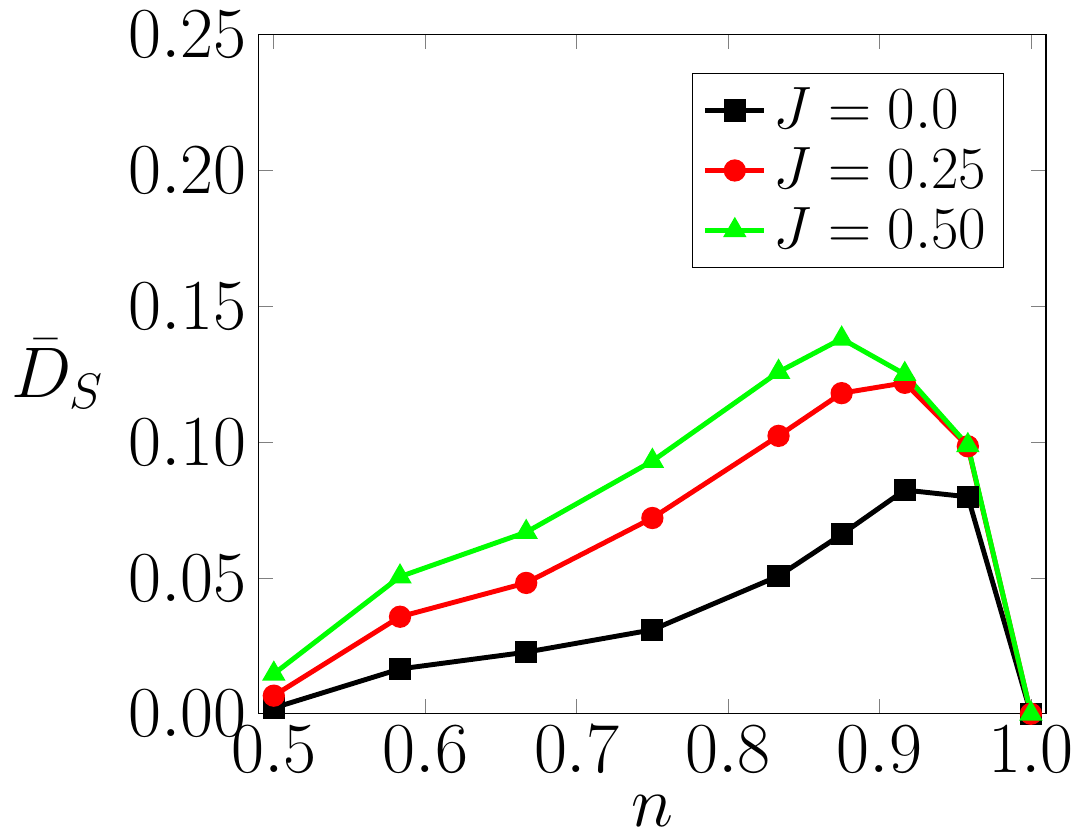}
\caption{(Color online) Low energy spin spectral weight defined in Eq.~\ref{eq:Ds}
as a function of electron filling for the values of $J$ indicated.} 
\label{fig:15}
\end{figure}

To better characterize quantitatively this spectral weight redistribution, 
we have evaluated the integrated low-energy spin spectral weight around the 
wave-vector transfer $(\pi,\pi)$ as a function of hole doping (see Fig.~\ref{fig:15}).
In particular, we have integrated the dynamical spin structure factor in the following 
rectangular region $k_x \in [k_F,\pi]$, $\omega \in [0,\Delta_S]$, 
where $\Delta_S$ is the spin gap at half-filling (in practice $\Delta_S$ is $0.125$, $0.25$ and
$0.375$, at $J=0.0$, $0.25$, and $0.50$, respectively), defining:
\begin{equation}\label{eq:Ds}
\bar{D}_S=\int_{k_F=\pi n}^{\pi} dk_x \int_{0}^{\Delta_S} S(k_x,\pi,\omega).
\end{equation}
By construction, the quantity above is zero at half-filling, because there is no weight below the spin gap
plus in this case $k_F$ reaches $\pi$. This corresponds to the intuitive
notion that, even if the binding energy for hole-pairs is finite in the half-filled case indicating
hole pair tendencies, if there are no holes in the system then there are no carriers 
that can generate robust pair-pair correlation functions. Only if there is a finite concentration of holes,
the magnetic mechanism can bind holes leading to a superconducting phase.
Thus, the quantity defined above is qualitatively consistent with expectations for pairing
in a doped magnetic system.

By hole doping, spin spectral weight is progressively transferred to 
the domain region of integration (which increases as $n$ decreases from 1) 
as a consequence of the commensurate-to-incommensurate effect 
observed, e.g., moving from $n=1$ to $n=0.9583$ in panels (a-b) of Fig.~\ref{fig:12}. 
The results in Fig.~\ref{fig:15} show that 
the low-energy spin spectral weight reaches 
a maximum at $n\simeq0.925$ for $J=0.0$, while  
the dominant peak slightly shifts to lower values as one increases the magnetic exchange 
interactions: $\bar{D}_S$ peaks at $n\simeq 0.9$ for $J=0.25$ and 
at $n\simeq0.875$ for $J=0.50$. For larger hole dopings, 
$\bar{D}_S$ decreases reaching is minimum for $n=0.5$. 

Results in Fig.~\ref{fig:15} clearly display similarities with the pairing strength studied 
in Fig.~\ref{fig:9}(b)
suggesting that the quantity $\bar{D}_S$ can be qualitatively used
as a measure of the pairing strength and it can be extracted experimentally 
from the magnetic excitation spectra.  Even the anomalous pairing-related ``bump'' 
at $n=0.5833$ in Fig.~\ref{fig:9}(b) also appears as a mild feature in Fig.~\ref{fig:15}.
These similarities should not be underestimated: Fig.~\ref{fig:15} was obtained
totally from magnetism, and independently of any pairing measurement in the ground state.

\section{Summary and Conclusions}\label{sec:conclusions}

In this publication, we have studied the pairing properties and 
magnetic excitation spectra of a generalized Hubbard model ($t-U-J$ model) on a 
ladder geometry. We have analyzed the behavior of the system by changing the 
magnetic exchange interaction $J$, the on-site Hubbard repulsion $U$, and the
electronic filling $n$.

With regards to ground state properties, our analysis confirms 
that the dominance of pairing correlations increases as the strength
of the superexchange interaction $J$ increases. For the case of $J=0.0$,
pairing is optimized for a Hubbard repulsion $U=6.0$ at the widely used
electronic filling $n=0.875$. Moreover, the pairing correlations strength
has an ``asymmetric superconducting dome'' shape with a broad maximum
around $5-10\%$ hole doping, remaining robust over a wide range
of hole densities until it becomes negligible at $50\%$ doping.

We have focused on providing a detailed analysis of the 
magnetic excitation spectrum and we searched
for connections with the pairing properties of the doped ground state.
With regards to the addition of the ``extra'' exchange interaction $J$, 
this new term mainly shifts the magnetic spectrum to higher energies, 
but qualitatively keeps unchanged the shape of the dispersive features. 
Interestingly, we noticed that by increasing $J$
spectral weight increases in the low energy region region around $(\pi,\pi)$.
Analogously, by increasing the 
on-site Hubbard repulsion, the spectral weight redistributes from high to 
intermediate-low energies, mantaining approximately the shape of
the main features.

More importantly, we have studied the properties of the dynamical spin 
structure factor as a function of hole doping. The results indicate that
the magnetic excitations dispersion along the line $(0,0)-(\pi,0)$ 
in the Brillouin zone, namely in the $k_{rung}=0$ branch of the spectrum,
do not change much varying hole doping in the interval $1 < n<0.8333$. 
On the other hand, the pairing correlation strength rapidly increases with doping
starting from a very small value at half-filling $n=1$, reaching a maximum 
around $n\simeq0.9$, and then further decreasing by hole doping to a negligible value
again at $n \sim 0.5$. At the same time, in the $k_{rung}=\pi$ branch
of the spectrum, a significant spectral 
weight redistribution away from the $(\pi,\pi)$ wave-vector transfer, 
characteristic of the half-filled case, is observed as a function of hole-doping. 
Low energy spin incommensurate features develop. Our results suggest that the
vicinity of $(\pi,\pi)$ is the portion of the spectrum that is related the
most with hole pairing, in agreement with a recent Quantum Monte Carlo study
supplemented by Maximum Entropy techniques of the two dimensional Hubbard model~\cite{huang2017decrease}. 

Even though obtained on ladders, our results are consistent with 
the general picture that emerged from recent RIXS and neutron scattering experiments
on two-dimensional cuprates, highlighting again 
the similarity of the physics of ladders and two-dimensional systems. Indeed, 
recent RIXS investigations of LSCO~\cite{dean2012spin,dean2013persistence,re:Wakimoto2015} 
have found the persistence of high energy magnetic excitations at the antiferromagnetic zone boundary 
from the underdoped up to the highly overdoped regime where superconductivity disappears. 
At the same time, neutron scattering experiments have shown that low energy magnetic excitations 
around the antiferromagnetic zone center are much reduced with doping~\cite{re:Wakimoto2007}.
In this work we find that, even for ladders, high energy 
magnetic excitations along the $k_{rung}=0$ branch 
do not change much up to large hole dopings, therefore appearing marginal 
to the pairing mechanism, while the main reason for the reduction of the pairing strength 
needs to be researched in the spectral 
weight at low energies around the antiferromagnetic zone center $(\pi,\pi)$.

In order to characterize quantitatively this spectral weight redistribution, 
we have evaluated the integrated low-energy spin spectral weight around the 
wave-vector transfer $(\pi,\pi)$ as a function of hole doping (see Fig.~\ref{fig:15} 
and Eq.~\ref{eq:Ds}). The qualitative similarity of the results shown in Fig.~\ref{fig:15} 
with the pairing strength extracted from a ground state analysis 
(Fig.~\ref{fig:9}(b)) suggest that the pairing correlation strength
$\bar{D}_S$ could be extracted 
experimentally directly from the magnetic excitation spectra.  
A recent neutron scattering study~\cite{PhysRevB.88.014504} 
of the spin gap evolution upon doping in 
the spin-ladder compound Sr$_{14-x}$Ca$_x$Cu$_{24}$O$_{41}$ 
has shown progress in the possibility of measuring the full 
spectrum response of strongly correlated ladders by changing the hole dopings. 
We urge neutron scattering experts to carry out inelastic neutron scattering experiments
on ladder materials over a wide range of doping to test our theoretical predictions.
Because of the clear similarities between cuprate ladders and layers, our conclusions
can tentatively be extended to two-dimensional systems as well.

\begin{acknowledgments}
The authors acknowledge useful conversations with Prof. C. Batista.
A.N. and E.D. were supported by the US Department
of Energy (DOE), Office of Basic Energy Sciences
(BES), Materials Sciences and Engineering Division. 
N.P. was supported by the National
Science Foundation Grant No. DMR-1404375. 
The work of G.A. was conducted at the Center for Nanophase
Materials Science, sponsored by the Scientic User Facilities
Division, BES, DOE, under contract with UT-Battelle.
Numerical simulations were performed at the Center for Nanophase Materials Sciences, 
which is a DOE Office of Science User Facility.
\end{acknowledgments}

\bibliography{thesis}

\end{document}